\documentclass{aa}
\usepackage{graphicx}

\usepackage[varg]{txfonts}
\usepackage{url}
\usepackage{color}
\usepackage{hyperref}

\newbox\grsign \setbox\grsign=\hbox{$>$} \newdimen\grdimen \grdimen=\ht\grsign
\newbox\simpropbox
\setbox\simpropbox=\hbox{\raise.5ex\hbox{$\propto$}\llap
 {\lower.5ex\hbox{$\sim$}}}\ht2=\grdimen\dp2=0pt

\newcommand{\msun}{{\rm M}_{\sun}}

\begin{document} 

\title{Doughnut strikes sandwich: the geometry of hot medium in accreting black hole X-ray binaries}
\titlerunning{Geometry of hot medium in accreting black hole binaries}

\author{Juri Poutanen \inst{1,2,3} \and Alexandra Veledina \inst{1,2,3} \and Andrzej A. Zdziarski \inst{4}}

\authorrunning{Poutanen et al.} 

\institute{
Tuorla Observatory, Department of Physics and Astronomy,  FI-20014 University of Turku, Finland
\\ 
\email{juri.poutanen@utu.fi, alexandra.veledina@utu.fi}
\and
Nordita, KTH Royal Institute of Technology and Stockholm University, Roslagstullsbacken 23, 10691 Stockholm, Sweden
\and
Space Research Institute of the Russian Academy of Sciences, Profsoyuznaya str. 84/32, 117997 Moscow, Russia 
\and
Nicolaus Copernicus Astronomical Center, Polish Academy of Sciences, Bartycka 18, PL-00-716 Warszawa, Poland \\
\email{aaz@camk.edu.pl}
}


\abstract{
We study the effects of the mutual interaction of  hot plasma and cold medium in black hole binaries in their hard spectral state on the value of the truncation radii of accretion discs. 
We consider a number of different geometries. 
In contrast to previous theoretical studies, we use a modern energy-conserving code for reflection and reprocessing from cold media. 
We show that a static corona above a disc extending to the innermost stable circular orbit produces spectra not compatible with those observed. 
They are either too soft or require a much higher disc ionization  than that observed. 
This conclusion confirms a number of previous findings, but disproves a recent study claiming an agreement of that model with observations. 
We show that the cold disc has to be truncated in order to agree with the observed spectral hardness. 
However,  a cold disc truncated at a large radius and replaced by a hot flow produces spectra which are too hard if the only source of seed photons for Comptonization is the accretion disc. 
Our favourable geometry is a truncated disc coexisting with a hot plasma either overlapping with the disc or containing some cold matter within it, also including seed photons arising from cyclo-synchrotron emission of hybrid electrons, i.e. containing both thermal and non-thermal parts. 
}
\keywords{
accretion, accretion discs -- black hole physics -- stars: individual: GX 339--4 -- X-rays: binaries -- X-rays: stars}

\maketitle

\section{Introduction}
\label{sect:intro}

Galactic black hole (BH) X-ray binaries (XRBs) are important cosmic laboratories which allow us to study many aspects of accretion. 
Thanks to their relative proximity, their X-ray brightness is often high, and thus their spectra and timing can be studied in detail. 
Given their size, their variability can be observed on a wide range of timescales, from the light travel time across the BH horizon to years. 
This allows us to observe their outbursts and state transitions between the two main spectral states, hard and soft. 

Most of the studies agree on the geometry in the soft state. 
The accretion flow consists of an optically thick and geometrically thin accretion disc \citep{SS73,NT73} extending down to the innermost stable circular orbit (ISCO). 
The blackbody-like spectrum of the disc is followed by a hard X-ray tail, which can have varying amplitude. 
They are probably due to coronal emission by hybrid, i.e. both thermal and non-thermal, electrons, which Compton scatter disc blackbody photons \citep[e.g.][]{PC98,gierlinski99,PV09}.

The hard-state data are well described by Comptonization proceeding in a hot medium. Location of this medium, above the accretion disc or within its truncation radius, has been a matter of debates for decades.
Related to this, there is an intense controversy regarding the value of the truncation radius. 
In transients, there is no inner disc during quiescence \citep{lasota96,DHL01}, so there has to be a truncated disc during initial parts of the outbursts. 
For many years, the dominant thinking was that the disc reaches the ISCO during the transition to the soft state. 
Lately, a number of publications have claimed the contrary (see the references below), namely that the disc is also close to the event horizon  in the hard state,  forming a sandwich-corona geometry. 
If this is the case, there are two questions that remain unanswered:  at what Eddington ratio does the disc reach the ISCO, and what are observational signatures of this phenomenon? 

\begin{figure}
\centerline{\includegraphics[width=\columnwidth]{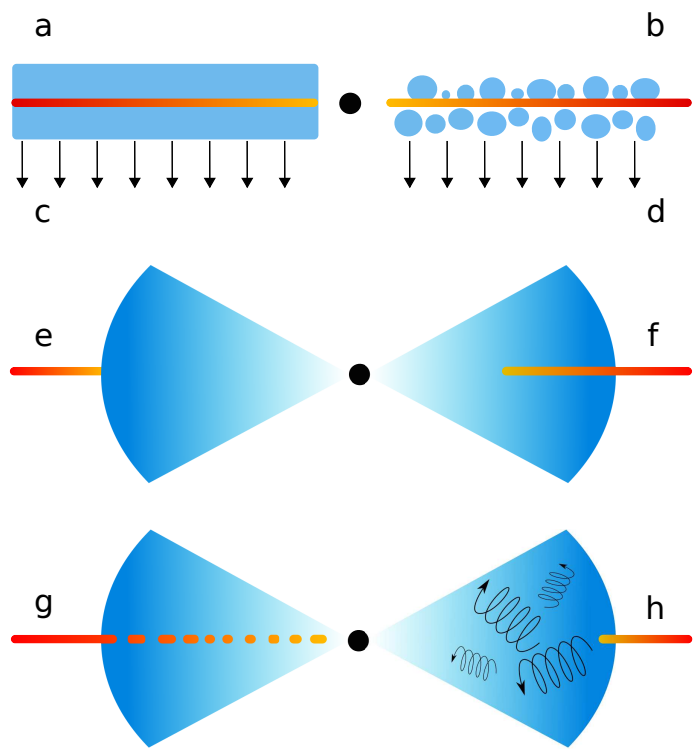}} 
\caption{Geometries of the central parts of the accretion flow proposed over time.
(a) Static sandwiching corona, (b) static patchy corona, (c) outflowing sandwiching corona, (d) outflowing patchy corona, (e) cold accretion disc detached from the hot flow and (f) intersecting with it, (g) hot flow with cold clumps, (h) truncated disc and hot flow with substantial cyclo-synchrotron radiation.
} \label{fig:geometry}
\end{figure}

Sandwich-corona models (see Fig.~\ref{fig:geometry}a) were proposed in \citet{HM91,HM93} in application to active galactic nuclei. 
These papers and many subsequent ones studied the disc-corona coupling in the limiting case of a passive disc with all the dissipation taking place in a corona above it  \citep[e.g.][]{HM91,HM93,SBS95,PS96,ZPM98}, and some treated the fraction of the dissipation in the corona above the disc as a free parameter \citep[e.g.][]{SZ94}. 
Models of self-consistent coronal emission and reflection were considered in \citet{SBS95}, \citet{PS96}, \citet{Sve96}, \citet{MBP01}, and \citet{MDM05}.
These models take into account up-scattering of reflected photons, but the intrinsic (viscous) dissipation in the disc was assumed to be zero.

A  lot of attention was paid to the energy balance between the disc and the corona. 
The emitted spectra were calculated self-consistently with the electron temperature in the corona, accounting for the feedback from the disc.
The radiation intercepted by the disc is partly reflected by Compton scattering in the upper layers, and is partly reprocessed and emitted at lower energies. 
The calculation of reflection in most of these models was done assuming neutral gas and neglecting line emission (the approach is described in \citealt{MZ95} and \citealt{PNS96}), and the reprocessed emission was approximated with a diluted blackbody.
These works showed that the models of the corona sandwiching the disc (slab-corona models) produce soft spectra, with photon index $\Gamma\gtrsim 2.1$ (see fig.~2 in \citealt{MBP01}).
The main reason for this is the large flux of the reprocessed photons which cool electrons in the corona. 
Specifically, since the corona emits about a half of its power towards the disc, there is an approximate equipartition between the power of the seed photons irradiating the corona and the power dissipated in the corona, which then leads to spectra roughly flat in $\nu F_\nu$, i.e. with $\Gamma\simeq 2$. 
Adding intrinsic dissipation then leads to even softer spectra \citep[e.g.][]{Sve96}.  
This is in contradiction to the observed hard-state spectra, with $\Gamma\sim 1.7$. 
Specifically, in the case of a hard-state spectrum of GX 339--4, \citet{ZPM98} found that the covering factor of the cold medium as seen from the hot plasma has to be $\lesssim$0.3 (see their fig.\,6), which rules out the geometry of a corona above the disc. 

A more advanced self-consistent slab-corona model was considered in \citet{MDM05}, who computed reprocessing in the disc in thermal and ionization equilibrium accounting for radiative transfer in lines using a two-stream approximation. 
They considered two models for the disc, one at  a constant density and one at a constant pressure of the cold phase, showing that in the first case the ionization parameter $\log \xi$ has to be at least 3.5 for the Comptonization spectrum to be consistent with the spectra observed in the hard state. 
On the other hand, they also showed that the Comptonization spectra for the second more physically realistic model of the constant pressure disc depend very weakly on the ionization parameter and cannot be reconciled with the data even for very high ionization. 
Regardless of the specific set-up, all the proposed models in this category  take into account the re-emission of the coronal radiation irradiating the disc as well as scattering of the reprocessed radiation in the corona. 

Harder spectra can be obtained in the patchy corona models (Fig.~\ref{fig:geometry}b), which can represent the magnetic flares above the accretion disc \citep{SBS95,Sve96}. 
In this case, most of the reprocessed photons directly escape towards the observer and are not intercepted by the coronal plasma, leading to the higher electron temperature and, subsequently, hard spectra. 
However, the luminosity of the soft reprocessed photons is similar to the luminosity escaping in the form of hard X-rays, again inconsistent with the hard-state data. 
This conclusion is drawn assuming zero flux from the viscous dissipation in the cold disc, which should be added to the soft reprocessed flux, increasing the soft flux even further. 
Another possibility is to assume that the corona is outflowing with mildly relativistic velocity \citep[Fig.~\ref{fig:geometry}c and d;][]{B99PE,MBP01}.
Because of the relativistic aberration, the coronal flux is beamed away from the disc, which reduces the reprocessed and the reflected components, making this model a feasible alternative for the hard-state spectra.

An alternative scenario considers the hot medium in the form of the hot flow surrounded by the cold disc truncated at large radius \citep[$r_{\rm in}\gtrsim 10R_{\rm g}$, where $R_{\rm g}\equiv GM/c^2$ is the gravitational radius, see Fig.~\ref{fig:geometry}e,][]{Esin97,Esin98,PKR97,Dove97}.
However, unlike the  sandwich corona, this geometry has the problem of seed photon starving, as only a small fraction of soft disc photons is intercepted by the hot flow. 
For example, if the disc truncation radius is $r_{\rm in} =100R_{\rm g}$, only about 0.1\% of the disc photons  enter the zone of major energy release (around $10R_{\rm g}$) as the result of reduction of the solid angle and Lambert law. 
Three solutions for this problem were proposed. 
In the first the disc overlaps with the hot medium \citep[Fig.~\ref{fig:geometry}f;][]{PKR97}.
The second option proposes additional soft photons from the cold medium in the form of clumps within the hot flow \citep{CFR92,ZPM98} or a residual accretion disc \citep{Liu07}, see Fig.~\ref{fig:geometry}g.
The third possibility is to consider cyclo-synchrotron radiation from the hot flow itself as the additional or even dominant source of seed photons \citep[Fig.~\ref{fig:geometry}h,][]{Esin97,WZ01,PV09,MB09,VPV13}.

Recently, the truncated disc/hot flow geometry has been questioned in a number of works.  
Papers claiming the accretion disc in the hard state of BH binaries extends to the ISCO, or almost to it, fall into two categories. 
In one category, the hard X-rays are supposed to be emitted by a very compact source located on the rotation axis of the BH (e.g. \citealt{parker15,fuerst15,miller15,walton16}). 
The way the compact source, known as the lamppost, emits, is not specified. 
If its emission is due to Compton  up-scattering of blackbody photons emitted by the disc, the size of the lamppost has to be relatively large in order to intercept enough photons \citep{DD16}, which is in conflict with most of the fitting results of this model. 
Also, the proximity of the lamppost to the horizon, found in many fitted models, gives rise to the problem of photon trapping of most of the emitted photons by the BH, and runaway e$^\pm$ pair production within the lamppost \citep{NZS16}, which  reduces the electron temperature to low values ($\sim$20~keV) inconsistent with the data. 

A  different argument is put forward by \citet{steiner17}, who  present a model for the accretion flow in the hard state attempting to self-consistently account for mutual interactions of the disc and the corona. 
In particular, they quantitatively take into account and stress the importance of Compton scattering by the corona of photons reflected from the disc. 
In order to compute the reprocessed emission that comes in the form of lines and recombination continua, they considered a constant density model of the disc.  
They also linked  the emission due to viscous dissipation within a disc truncated at an inner radius, $r_{\rm in}$, to the coronal emission by taking into account conservation of photons.  
They concluded that for the hard state a large truncation radius, $r_{\rm in}/R_{\rm g}\gg 1$, would require very high values, even super-Eddington values,  of the accretion rate. 
This would argue strongly against the correctness of hard-state fit results obtaining a large $r_{\rm in}$ in luminous hard states, in particular those for the low-mass X-ray binary GX 339--4 \citep{BZ16}. 
We note that the \citet{steiner17} approach relies on the assumption that  most of the seed photons  come from the viscous dissipation in the accretion disc, while the amount of soft photons coming from reprocessing was heavily underestimated, and other sources of seed photons in the hard state,  for example from  the synchrotron emission, were ignored (see Sect.~\ref{sec:comp_sandwich} for details).
A major shortcoming of  this work is that the cooling of the electrons in the corona by the reprocessed emission coming from the disc and the resulting coronal spectrum were not computed self-consistently.

In this paper, we reconsider the issue of energy balance between the disc and the corona. 
We calculate the self-consistent Comptonization spectra of the hot medium in various geometries and convolve them with the most recent reflection model {\sc xillver} \citep{GK10,GK13}.  
We show that the sandwich-corona geometry is not a viable alternative for the hard state, even accounting for the Comptonization of the ionized reflection emission, the main reason being the substantial soft photon flux from the reprocessed emission, capable of cooling the coronal medium and leading to too soft spectra. 
We compare our results to previous works involving the geometry of sandwiching corona and revisit the arguments against the truncated disc geometry. 
We discuss different accretion flow geometries in the context of recent multiwavelength discoveries and conclude that the truncated disc/hot flow geometry with cyclo-synchrotron seed photons is the only scenario capable of explaining the whole complex of the  hard-state data.

\section{Self-consistent modelling of the hot medium in various geometries}
\label{sect:modelling}

\subsection{Sandwich corona}
\label{sec:sandwich}

\begin{figure}
\centerline{\includegraphics[width=0.95\columnwidth]{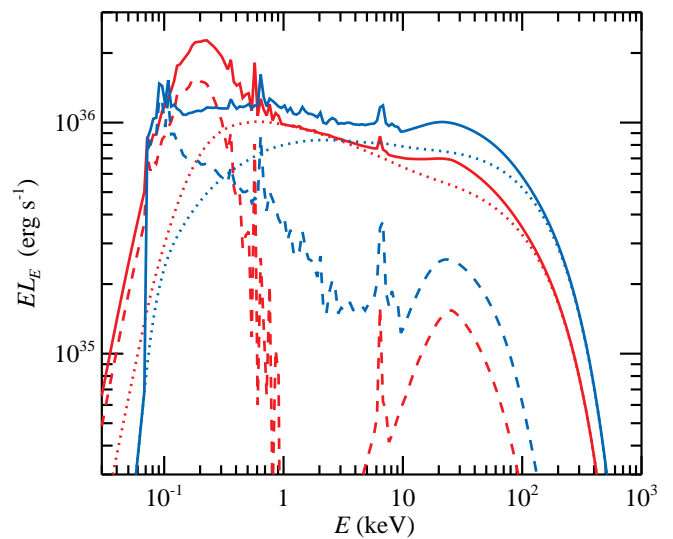}}
\caption{Self-consistent spectra from a homogeneous slab corona of $\tau_{\rm T}=0.4$ above a disc  as observed at inclination $i=60\degr$. 
The red and blue curves correspond to $\log \xi=1$ and $3$, respectively. 
The dotted lines show the Comptonization continuum, the dashed lines show the reflection {\sc xillverCp} spectrum together with reprocessed blackbody emission, and the solid lines give the total spectrum. 
The reflection spectra have been re-binned at a resolution $\Delta \log E=0.025$. 
The bolometric luminosity of the reflection spectrum here is smaller than that of the  Comptonization continuum because part of the reflected/reprocessed emission is Compton scattered in the corona.
} \label{fig:slab}
\end{figure}

We  discuss here  the simplest model of the slab corona above a cold disc. 
The exact solution for this problem was first obtained by \citet{SBS95}, \citet{PS96}, and \citet{MBP01} assuming reflection from neutral material \citep{MZ95,PNS96}. 
Here we use the newer reflection model, {\sc xillverCp}\footnote{\url{http://www.sternwarte.uni-erlangen.de/~dauser/research/relxill/}} 
\citep{GK10,GK13,Dauser16} in the form of precomputed tables.\footnote{We note that {\sc xillverCp} table model contain errors, in particular, all spectra at one particular inclination of $\cos i=0.65$ are in error and we interpolated the spectra to this angle using the results for the neighbouring angles. This error may cause the fits with {\sc xillverCp} or its relativistically smeared analog {\sc relxillCp} to be stacked close to this specific inclination leading to unrealistically small errors on the inclination. }  
In this model, the reflection spectra from a constant density slab are computed for the intrinsic Comptonization continuum described by the {\sc nthcomp} model \citep{ZJM96} incorporated to the {\sc xspec} package \citep{Arn96}. 
The main parameters of {\sc nthcomp} are the photon spectral index, $\Gamma$, and the electron temperature, $kT_{\rm e}$. 
The blackbody seed photons are assumed to have a temperature $kT_{\rm seed}$ of 50 eV. 
The reflection spectrum depends on the ionization state of the material, described by the parameter $\xi=4\pi F_{\rm X}/n$, where $F_{\rm X}$ is the integrated flux over the incident spectrum in the range 1--$10^3$ Ry and $n$ (cm$^{-3}$) is the gas density (particle concentration). 
It also depends on the metal abundance, which we assume here to be solar. 
The normalization of the reflection spectrum is given in the Appendix of \citet{Dauser16}.
The reflection spectrum contains the continuum reflected by coherent electron scattering as well as lines and recombination continua. 
Because the Klein--Nishina effect is ignored in the calculation of the reflection, the shape of the cut-off of the reflected spectrum energies above 50 keV is not correct (compare spectra from fig.~2 in \citealt{GK13} with the spectra presented in \citealt{MZ95,PNS96}; see also fig.~1 in \citealt{NZS16}). 
This may introduce an error of the order of 10--20\% to the reflection albedo, $a$, and the reprocessing efficiency. 
The reflection spectra in the {\sc xillverCp} model extend down to only 70 eV. 
This  sometimes results in a significant difference between the angle-integrated reflected and illuminating fluxes  which is inconsistent with the energy conservation. 
We solved this problem by adding  a soft component with the blackbody-like spectrum of temperature 50 eV to the reflected spectrum. 
Thus, the emission from the cold disc  consists of two components: the reflection  of the coronal X-rays and an additional soft component  (see red dashed curve in Fig.~\ref{fig:slab}). 

Electrons  in the corona Comptonize disc emission into a power-law-like  spectrum, which turns over at energies slightly above $kT_{\rm e}$ \citep{SBS95}. 
The main parameter defining the Comptonization spectrum is the Thomson optical depth, $\tau_{\rm T}$, of the corona. 
The coronal electron temperature is not a model parameter, but is computed self-consistently by solving for the energy balance \citep{PS96}. 
The angle-dependent Comptonization spectrum in a slab geometry is computed using the code described in \citet{PS96}, which utilizes an exact Compton scattering redistribution function.
The code is similar to the {\sc compps} model used in the {\sc xspec} package,  the only difference being that reflected photons are now scattered in the corona. 
To evaluate the reflection spectrum, we first compute the angle-averaged Comptonized spectrum of radiation going from the corona down towards the disc. 
This spectrum is then fit in the 2--10 keV range by a power law to get the photon index $\Gamma$, which together with the current value of $kT_{\rm e}$ is used to obtain the angle-dependent reflection spectrum by interpolation from the precomputed tables. 
We note that the angle-averaged Comptonization spectra computed using {\sc nthcomp} and {\sc compps} are very close for $kT_{\rm e}\lesssim 100$ keV \citep{ZJM96,ZLGR03}. 
This justifies  using  the reflection tables for {\sc nthcomp}. 
We  then compute the reflected flux in the 0.07--$10^3$ keV range (where the {\sc xillverCp} model is computed). 
The  remaining fraction of the bolometric Comptonized downward-directed flux is assumed to be thermalized in the disc and emitted as a blackbody of the temperature of 50 eV. 
This guarantees the correct energy balance in the disc. 
The self-consistent solution for the corona, which requires that the flux emitted by the disc is equal to the Comptonized flux directed towards the disc, is obtained by iterating the coronal temperature. 
Thus, we obtain the equilibrium $kT_{\rm e}$ and the escaping spectrum. 

 \begin{figure}
\centerline{\includegraphics[width=0.95\columnwidth]{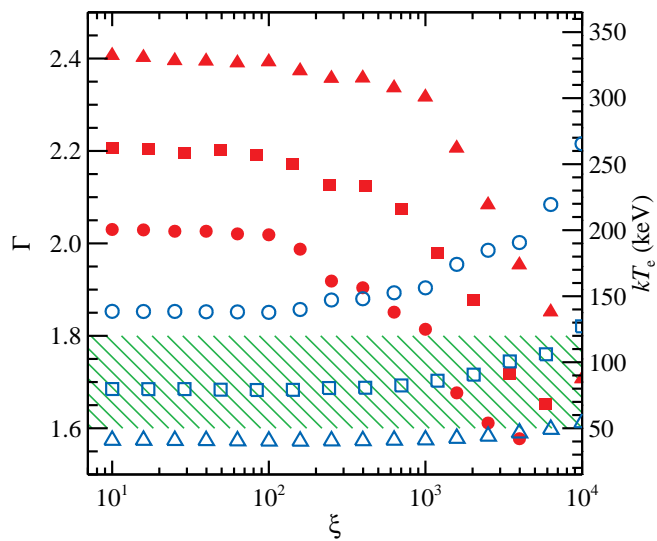}}
\caption{Dependence of the 2--10 keV photon spectral index $\Gamma$ of the Comptonized spectra from the sandwich corona (red filled symbols and left axis) and of the coronal temperature $kT_{\rm e}$ (blue open symbols and right axis)  on the ionization parameter for various coronal optical depths. 
Circles, squares, and triangles correspond to $\tau_{\rm T}=0.2,\,0.4$, and 0.8, respectively. 
The shaded area represents the typical observed range for $\Gamma$ (1.6--1.8) and $kT_{\rm e}$ (50--120~keV) \citep{wardzinski02,ZG04,IPG05,Torii11,BGS17}.
} \label{fig:slab_gamma_te}
\end{figure}

Examples of the resulting spectra from the slab corona as observed at some typical inclination of $i=60\degr$ are shown in Fig.~\ref{fig:slab} for two values of $\log \xi = 1$ and 3 and a fixed Thomson optical depth of $\tau_{\rm T}=0.4$. 
This value of $\tau_{\rm T}$ gives the equilibrium coronal temperature of $kT_{\rm e}=80$--100~keV (see blue open squares in Fig.~\ref{fig:slab_gamma_te}), which is typical for the BH XRBs in the hard state \citep{wardzinski02,ZG04,IPG05,Torii11,BGS17}.
We see that in a weakly ionized case with $\log \xi<2$, the Comptonized spectrum is rather soft (see the red squares in Fig.~\ref{fig:slab_gamma_te}) with $\Gamma\approx 2.2$ (slightly depending on the inclination $i$), which is consistent with previous results obtained for neutral reflection (see e.g. the high compactness range of the slab model in fig.~6 in \citealt{Sve96} or the $h/r=0$ case in fig.~2 in \citealt{MBP01}). 
The reflection amplitude in the total spectrum is rather small here because only a fraction of the reflected flux, $\exp(-\tau_{\rm T}/\cos i)$, is transmitted through the corona without scattering. 
At the intermediate ionization of $\log \xi=3$, the coronal spectrum becomes harder with $\Gamma\approx 2.04$. 
In both cases the spectrum is much softer than those observed in the hard state of BH XRBs with $\Gamma\approx$1.6--1.8.
We also note that the Comptonized spectra computed using the {\sc xillverCp} reflection model are indistinguishable from those computed using the simpler reflection model {\sc pexriv} for $\log \xi<2$; only at high ionization will  the presence of lines and recombination continua in the {\sc xillverCp} model  start to make a difference (see Sect.~\ref{sec:comp_refl}). 
However,  the resulting coronal spectrum is soft with $\Gamma>2$. 

The situation changes at high ionizations with $\log\xi\geq 3.5$ (see Fig.~\ref{fig:slab_gamma_te}). 
Now the disc is strongly ionized and  reprocesses very little radiation; most of the X-rays are scattered back. 
For $\log\xi=3.5$, the corona becomes photon-starved resulting in a harder spectrum with $\Gamma\approx 1.75$, with no visible spectral features seen in the {\sc xillverCp} spectrum. 
At even higher ionization the disc acts like a mirror resulting in the very hard coronal spectrum. 
Thus, we find that the slab corona can produce spectra similar to the hard states of BH XRBs only for very high ionizations $\log\xi\gtrsim 3.5$ (see also fig.~5 in \citealt{MDM05}),  the lower boundary coinciding with the upper limit of the observed ionization levels for sandwich corona fits $2.0\lesssim\log\xi\lesssim 3.4$ \citep{parker15,BZ16,steiner17,BZP17}.

Higher coronal optical depth, $\tau_{\rm T}=0.8$, results in even softer spectra (see red triangles in Fig.~\ref{fig:slab_gamma_te}) and predicts very low coronal temperatures of below 50 keV (blue open triangles), not consistent with the data. 
The main reason for spectral softening is that at higher $\tau_{\rm T}$ a larger fraction of disc radiation would be scattered back to the disc. 
For the same spectral slope this would mean that more radiation is striking the disc than is reflected, which breaks the energy balance. 
Thus, the corona  has to adjust by cooling and producing a softer spectrum.
For the same reason a lower optical depth, $\tau_{\rm T}=0.2$, gives slightly harder spectra with $\Gamma=2.03$ at low $\xi$. 
Harder spectra with $\Gamma<1.8$ can be achieved at $\log\xi>3$. 
However, the coronal temperature in this case is in excess of 150 keV (blue open circles), which is well above the observed temperature values in most of the observations \citep{wardzinski02,ZG04,IPG05,Torii11,BGS17}.

The observed range of $\Gamma$ (1.6--1.8) and $kT_{\rm e}$ (50--120~keV) is shown as a shaded area in Fig.~\ref{fig:slab_gamma_te}.
We see that most of the models demonstrate either too soft spectra or too high electron temperatures.
Hence, the slab-corona model with no intrinsic disc dissipation requires much fine-tuning in both optical depth, $\tau_{\rm T}=0.4\pm 0.1$, and ionization parameter, $\log\xi>3.5$, to be consistent with the observed slope of the Comptonized continuum. 
Furthermore,  in the hard state the best-fitting ionization parameter is usually much smaller (e.g. \citealt{Plant14,parker15,BZ16,steiner17,BZP17}).

We note here that the Compton reflection is mirror-like only in the Thomson limit, i.e. for spectral cut-off below $E\approx 50$ keV. 
In reality, the Klein--Nishina effect results in the absorption of a fraction of the energy of incident photons at $E\gtrsim 50$ keV (see fig.~11 in \citealt{ZLGR03}).
This effect, not treated in {\sc xillverCp}, leads to additional cooling and softening of the observed spectrum. 
For example, in the case of $kT_{\rm e}=120$~keV and $\Gamma\simeq 1.5$ considered in \citet{ZLGR03}, the reflection albedo is still as low as 0.48, which means that 52\% of the irradiating luminosity is re-emitted at low energies and serve as seed photons, cooling the plasma. 
Thus, the spectral indices computed  using {\sc xillverCp} reflection are the lower limits and the actual spectra should be softer. 
Furthermore, the approximation of the constant-density slab used by {\sc xillverCp} predicts too few soft photons compared to a more physically realistic case of reflection from a constant pressure slab \citep[fig.~5 in][]{MDM05} or a slab in hydrostatic equilibrium \citep[fig.~8 in][]{NKK00}. 
The latter models  would thus produce softer Comptonization coronal spectra and would  rule out the slab-corona model for the hard state completely \citep[see also][]{ND01}.

\subsection{Hot flow with cold clumps or overlapping with the disc}
\label{hot_flow}

To  make the spectra harder than typically predicted by the slab corona, we need to somehow reduce the amount of soft photons coming from the disc. 
It is straightforward to assume that the disc does not reside under the hot medium all the way down to the black hole, but that it is truncated at some radius.
Self-consistent models for the central hot corona--approximated as either a sphere of  radius  $r_{\rm c}$ irradiating an outer disc with some viscous dissipation (or without) and truncated at some radius $r_{\rm in}$ (allowed to overlap with the corona) or as a hot flow with embedded clouds--have been considered by \citet{ZPM98,ZLS99}, and see also \citet{PKR97}. 
 In this model, the corona Compton-scatters blackbody and reflected photons emitted by either the outer disc or cold clouds, reaching the luminosity
\begin{equation}
L_{\rm C}=f_{\rm sc} A(\Gamma, T_{\rm bb}, T_{\rm e}) L_{\rm bb},
\label{LC}
\end{equation}
where $f_{\rm sc}$ is the fraction of the blackbody photons that irradiate the corona (which depends on the geometry) and $A(\Gamma, T_{\rm bb}, T_{\rm e})$ is the Comptonization amplification factor (which depends on the spectral index, $\Gamma$; the blackbody temperature, $T_{\rm bb}$; and the electron temperature, $T_{\rm e}$). 
The blackbody luminosity in turn originates from reprocessing of the coronal Compton-scattered photons irradiating the cold medium and the intrinsic viscous dissipation in the disc: 
\begin{equation} \label{lbb_irr}
L_{\rm bb}=L_{\rm intr}+L_{\rm irr},
\quad
L_{\rm irr} = R_{\rm F}(1-a)L_{\rm C}.
\end{equation}
The reflection fraction $R_{\rm F}$ follows from the assumed geometry. 
The albedo, $a$, depends on the incident spectrum and the ionization properties of the medium defined by the ionization parameter $\xi$.
For a given geometry, the model determines the X-ray spectral index, $\Gamma$, and the strength of reflection, which are found to be correlated, as observed  (see e.g. \citealt{ZLGR03,Gilfanov10}). 

\begin{figure}
\centerline{\includegraphics[width=0.95\columnwidth]{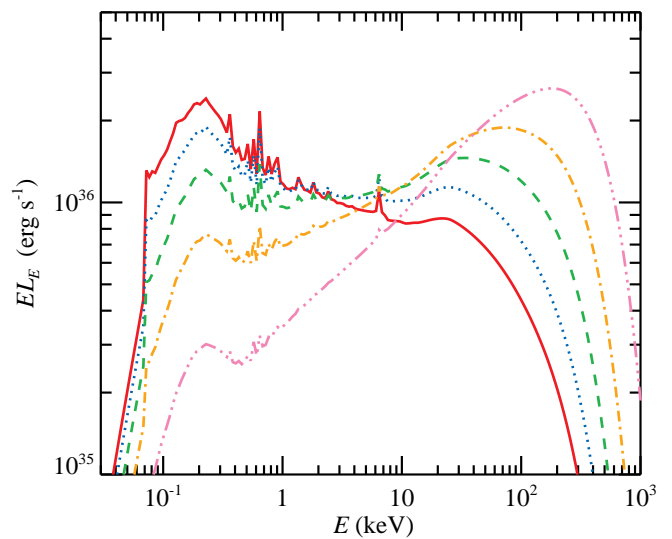}}
\caption{Self-consistent spectra from hot thermal plasmas mutually interacting with cold media without internal dissipation.
The Thomson optical depth of the hot flow down to the central plane is $\tau_{\rm T}=0.4$ and the spectra are observed at inclination of $i=60\degr$. 
Red solid, blue dotted, green dashed, brown dot-dashed and pink triple-dot-dashed lines correspond to the covering fraction of the cold clumps of $f_{\rm cl}=1.0,\, 0.8,\, 0.6,\, 0.4$ and $0.2$, respectively. 
The ionization parameter here is $\log\xi=1$. 
All spectra correspond to the same bolometric luminosity. 
} \label{fig:clumps}
\end{figure}

To illustrate the model, we consider a hot electron slab with the central plane filled by cold clouds (i.e. $f_{\rm sc}=1$ in this case) with the covering factor $f_{\rm cl}$, with $f_{\rm cl}=1$ corresponding to the slab-corona case described in Sect.~\ref{sec:sandwich}. 
As in the case of the sandwich-corona geometry, we inject blackbody and reflected photons from the cloud surface. 
 
They are then Comptonized by the hot medium described by $\tau_{\rm T}$ (now measured from the central plane to the outer surface) and the electron temperature, $kT_{\rm e}$. 
We compute the Comptonization angle-dependent spectrum and the reflection exactly as in Sect.~\ref{sec:sandwich} with {\sc compps} and {\sc xillverCp} models, respectively. 
Because  $f_{\rm cl}$ can now be below unity, only a fraction of the Comptonizing radiation crossing the central plane produces blackbody and reflected photons; the rest  pass to the other side of the hot slab (which we simulate by mirror reflection). 
The value of $f_{\rm cl}$ close to zero means that there is very little reprocessing and very few seed soft photons from the clouds cooling the hot flow. 
Similarly to the slab corona, we search for the equilibrium coronal electron temperature by iterations requiring that the Comptonized coronal flux through the central plane times the factor $f_{\rm cl}$ equals the flux emitted by the cold clouds.

\begin{figure}
\centerline{\includegraphics[width=0.95\columnwidth]{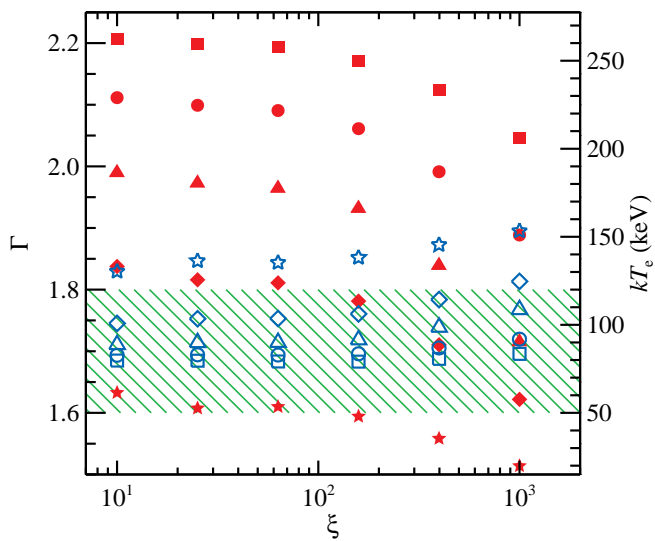}}
\caption{Same as Fig.~\ref{fig:slab_gamma_te}, but for the hot flow with cold clumps. 
Here $\tau_{\rm T}=0.4$ and squares, circles, triangles, diamonds and stars correspond to cold clump covering factors $f_{\rm cl}=1.0,\, 0.8,\, 0.6,\, 0.4$ and $0.2$, respectively. 
} \label{fig:clumps_gamma_te}
\end{figure}

Figure~\ref{fig:clumps} shows example spectra in the case of hot flow with embedded cold clouds. 
The decreasing covering factor leads to the hardening of the spectrum because the same power has to be emitted by the flow with far fewer soft seed photons. 
This is possible only if the coronal temperature is higher (i.e. the Comptonization spectrum is harder) as demonstrated in Fig.~\ref{fig:clumps_gamma_te}. 
We also note that the reflection spectrum and, in particular, the line strength vary with $f_{\rm cl}$ because the softer spectra correspond to effectively greater ionization \citep[see Sect. 3.3 in][]{GK13}. 
In order to reproduce the hard-state spectra of GX 339--4 with $\Gamma\approx 1.8$, the factor $f_{\rm cl}\sim 0.4$ is required, which is consistent with the results of \citet{ZPM98}. 

Figure~\ref{fig:clumps_gamma_te} shows the dependence of the spectral index and the coronal temperature on the ionization parameter for different values of $f_{\rm cl}$. 
The general trend is the following: smaller $f_{\rm cl}$ leads to harder spectrum: it evolves from $\Gamma\approx 2.2$ for $f_{\rm cl}=1$ to $\Gamma\approx 1.6$ for $f_{\rm cl}=0.2$. 
The coronal temperature  correspondingly increases from 80 keV to 140 keV. 
As in the case of the slab corona, increasing ionization leads to a higher reflection albedo and a smaller soft photon luminosity, which result in hardening of the spectra. 
Thus, as $\log\xi$ changes from 1 to 3, $\Gamma$ decreases by 0.2. 
We see  in the context of this model that in order to reproduce the observed spectral indices $\Gamma\approx 1.7\pm0.1$ in the hard state, we either need to have rather small $f_{\rm cl}<0.4$ or very high ionization with $\log\xi>3$. 
As in the slab-corona model, high $\xi$ is unlikely to be a solution because no high-ionization spectral features are seen in the hard-state data \citep{Plant14,parker15,BZ16,steiner17,BZP17}. 
Thus, the only physically realistic solution is to assume small $f_{\rm cl}$. 
However, the reflection amplitude in this case, $\propto\! f_{\rm cl}\exp(-\tau_{\rm T}/\cos i)$, is much smaller than the typically observed hard-state reflection fraction of $\gtrsim 0.2$ (see e.g. \citealt{parker15,fuerst15,BZ16,BZP17} for recent studies) requiring an additional source of reflection, likely from the truncated disc. 
 
The case of a geometrically thick hot flow overlapping with an outer disc is similar, except that there is a significant fraction of blackbody photons that are not scattered in the hot flow, but escape to the observer directly. 
Thus, the relative amplitude of the soft component is higher.

 \subsection{Patchy and outflowing corona models}
\label{sec:outflow}

As  we showed above, a reduction in the reprocessing efficiency of geometries where the hot flow is surrounded by the cold disc and in the hot flow with cold clumps  produces hard spectra. 
A similar reduction can be achieved in patchy corona models where the hot phase is situated in localized regions (flares) above the cold accretion disc \citep{GRV79,HMG94,SBS95,Sve96}. 
In this case, the higher  the flaring region above the disc, the smaller  the  fraction of reprocessed photons that come back to the flare. 
This reduces the cooling and produces harder spectra. 
This model, however, predicts an anticorrelation between the amplitude of reflection $R$ and spectral index $\Gamma$ in contradiction with the observations \citep{ZLGR03}.  
Furthermore, as the flares emit nearly isotropically, the reprocessed blackbody component from the accretion disc is expected to be of a comparable luminosity to the observed Comptonized emission from the flare, also contradicting the data, which show that the soft component is an order of magnitude less luminous  \citep[e.g.][]{FZA01,KVT16}. 

Alternatively, the dynamical outflowing corona  \citep{B99PE,MBP01} can satisfy the observational spectral constraints. 
In this model, the higher the velocity of the flaring material away from the disc, the more the radiation is beamed away from the disc and the lower the reprocessing efficiency leading to harder spectra. 
This model can reproduce the observed $R$--$\Gamma$ correlation  \citep{ZLS99,ZLGR03,GCR99,RGC01,Gilfanov10}  and is consistent with the observed range of spectral indices and electron temperatures.
However, it requires  the velocity to decrease on  transition from the hard to the soft state;  the reasons for this are unknown. 

Furthermore, this model heavily relies on the assumption that the energy dissipation occurs exclusively in the corona with no intrinsic disc dissipation. 
The intrinsic disc radiation in this model will significantly influence the resulting spectra, because the energy density of the disc radiation as observed in the comoving frame of the plasma will actually be higher than that measured in the disc frame relevant for the static corona models (see Sect.~\ref{sec:intrinsic} below). 
Thus, the model also requires that the location of the energy release should change from the cold disc in the soft state to the corona in the hard state for  reasons that are not understood or discussed.

\subsection{Intrinsic disc dissipation}
\label{sec:intrinsic}

The models discussed above assumed that all the energy is dissipated in the hot phase (corona or hot flow) and there is no intrinsic viscous dissipation happening within the disc or cold clouds, which is not realistic. 
An additional dissipation of energy in the dense matter leads to an increase in the flux of soft photons, in particular those irradiating the hot plasma. 
This will then lead to a decrease in the electron temperature and the associated softening of the scattered spectra (at a given $\tau_{\rm T}$). 

In the framework of the model of the hot flow overlapping with an outer disc \citep[see e.g.][]{ZLS99}, an intrinsic dissipation in the disc was considered by \citet{IPG05}, who also took into account the decrease in the strength of the observed Compton-reflection feature due to scattering in the corona. 
For completeness, here we also  calculate the Comptonization spectra from the slab-corona model and the hot flow with cold clumps model  taking into account the intrinsic disc dissipation. 

\begin{figure}
\centerline{\includegraphics[width=0.95\columnwidth]{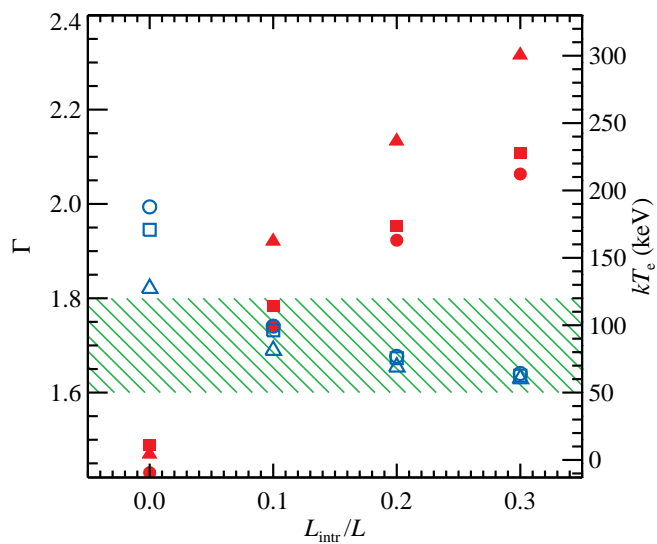}}
\caption{Same as Fig.~\ref{fig:slab_gamma_te}, but the photon index and the equilibrium electron temperature of the hot phase are shown as a function of the ratio of intrinsic disc dissipation rate to the total luminosity. 
Squares show the case of $\log\xi=1, f_{\rm cl}=0.1$; 
circles  $\log\xi=3, f_{\rm cl}=0.1$; 
and triangles  $\log\xi=4, f_{\rm cl}=1.0$. 
In all cases $\tau_{\rm T}=0.4$.  
} \label{fig:intr_disc}
\end{figure}

The main new parameter in these models is the ratio of the luminosity associated with the intrinsic disc dissipation $L_{\rm intr}$ to that of the total luminosity.  
The intrinsic disc spectrum is assumed to have the blackbody shape of temperature 50 eV. 
In Fig.~\ref{fig:intr_disc}, we show the spectral slopes and the equilibrium electron temperatures as functions of this ratio for three different pairs of ionization parameter and cold clump covering  factor. 
For large ionization parameter of $\log\xi=4$ in a slab-corona (sandwich) model ($f_{\rm cl}=1.0$), the intrinsic disc dissipation rate cannot exceed 10 \% of the total luminosity, otherwise the Comptonization spectrum becomes too soft.
In the case of the hot flow with cold clumps of covering factor $f_{\rm cl}=0.1$, we get a similar constraint on the intrinsic disc luminosity nearly independently of the ionization parameter (compare squares and triangles in Fig.~\ref{fig:intr_disc}). 
Thus, it is clear that in order to be consistent with the observed range of spectral slopes and electron temperatures, for the considered models where all seed soft photons propagate through  the hot phase, the intrinsic disc dissipation cannot exceed 10 \% of the total luminosity.   
Additionally, as shown before, for the sandwich-corona model the condition of $\log\xi>3.5$ has to be satisfied, while for the hot flow with cold clumps model either $\log\xi>3.5$ or $f_{\rm cl}<0.4$ are required.

The intrinsic disc dissipation also affects the spectra from the  outflowing corona discussed in Sect.~\ref{sec:outflow}. 
Relativistic aberration effects increase the solid angle of the disc seen from the outflowing medium (up to a factor of 2 for an infinitely flat disc and an ultrarelativistic outflow velocity).
In addition, the photon spectrum and total energy are modified.
The total spectrum thus softens in the case of no dissipation, similar to the case of zero outflow velocity (sandwich corona).
The strong dependence of the resulting spectral slope on the dissipation fraction is complemented by the dependence on the outflow velocity \citep[see e.g. fig.~8 in][]{MBP01}, requiring some fine-tuning of parameters to place the modelled spectral slopes in the observed range.

\subsection{Hot flow with cyclo-synchrotron radiation}
\label{synchrotron}

\begin{figure}
\centerline{\includegraphics[width=0.95\columnwidth]{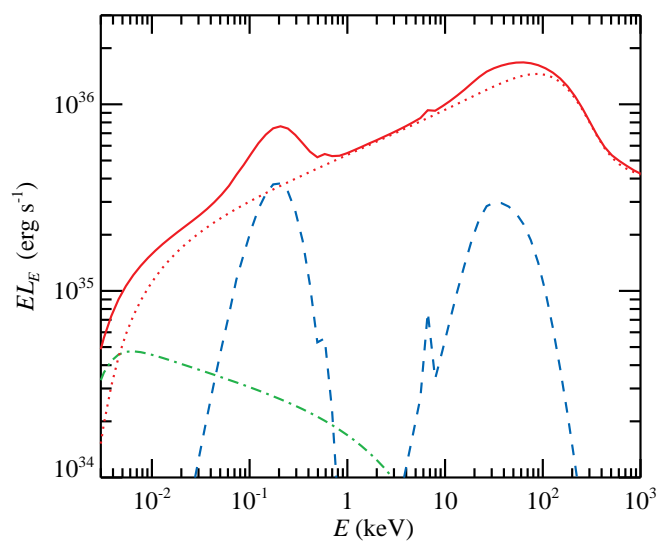}}
\caption{Example of a synchrotron self-Compton spectrum from hot plasma with a hybrid electron distribution. 
The components due to reflection/reprocessing with the reflection fraction of 0.3 and the ionization parameter of $\log\xi=1$ and the blackbody of the temperature of 50 eV due to the reprocesses emission are shown together by the blue dashed curve. 
The escaping synchrotron spectrum is shown by the green dot-dashed line, synchrotron Comptonization spectrum is shown by the red dotted curve, and the total spectrum is given by the red solid curve. 
} \label{fig:hybrid}
\end{figure}

The above discussion concerns models in which the disc blackbody is the sole source of seed photons for Comptonization. 
On the other hand, an additional source is provided by the cyclo-synchrotron process, which is likely to be copious given that magnetic fields are required in accretion flows for an outward transfer of the angular momentum. 
Various types of radiatively inefficient accretion models rely exclusively on partially self-absorbed 
cyclo-synchrotron photons as seeds for Comptonization, in particular the advection-dominated accretion flow (ADAF) models, starting with \citet{NY94}, and followed by a very large number of works (see \citet{YN14} for a review). 

These  models assume that the electron distribution in the hot flow is Maxwellian. 
This results in a relatively low rate of the synchrotron emission at low temperatures, $kT_{\rm e}\lesssim 100$ keV \citep{WZ00}. 
Low synchrotron luminosity in turn results in inefficient electron cooling (in the absence of other sources, such as cold accretion disc), and the electron temperature increases  to the point where the radiative losses become comparable to characteristic heating rates, which for ADAF models is typically achieved at $kT_{\rm e}\gtrsim 200$~keV \citep[e.g.][]{yuanaaz04,YZ07}.
 However, the electron distribution is very likely to have a non-thermal tail (i.e. to be hybrid; \citealt{PC98,Coppi99,gierlinski99}) due to the energy release proceeding in the form of magnetic reconnection \citep[e.g.][]{SS14,Bel17}.
This is supported by the observations of MeV emission in both hard and soft states of black holes \citep{McConnell94,McConnell02,Ling97,Grove98,wardzinski02,DBM10,ZMC17}. 
The presence of the high-energy electron tail then significantly enhances the synchrotron emission \citep{WZ01}.  

\begin{figure*}
\centerline{
\includegraphics[width=0.92\columnwidth]{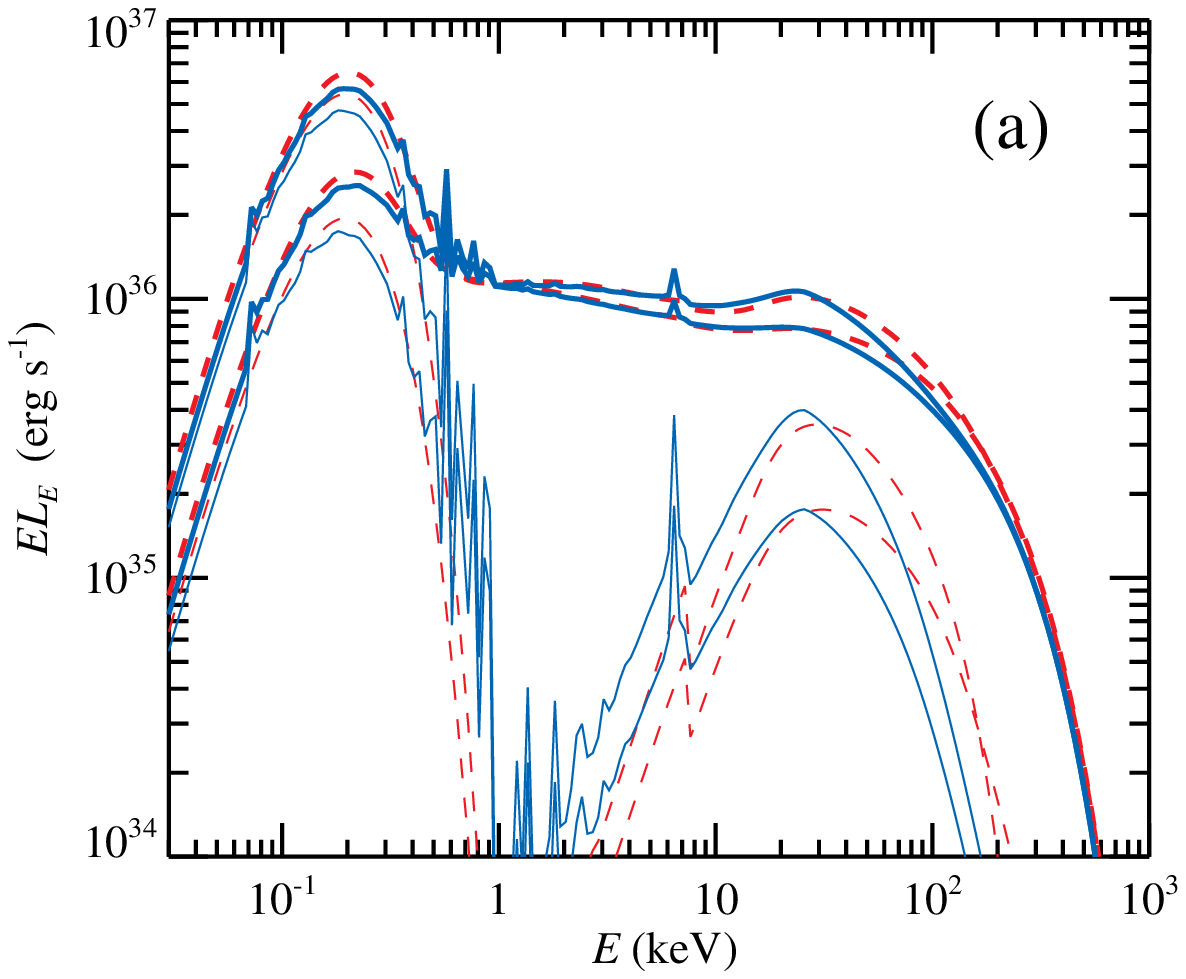}
\hspace{0.5cm}
\includegraphics[width=0.92\columnwidth]{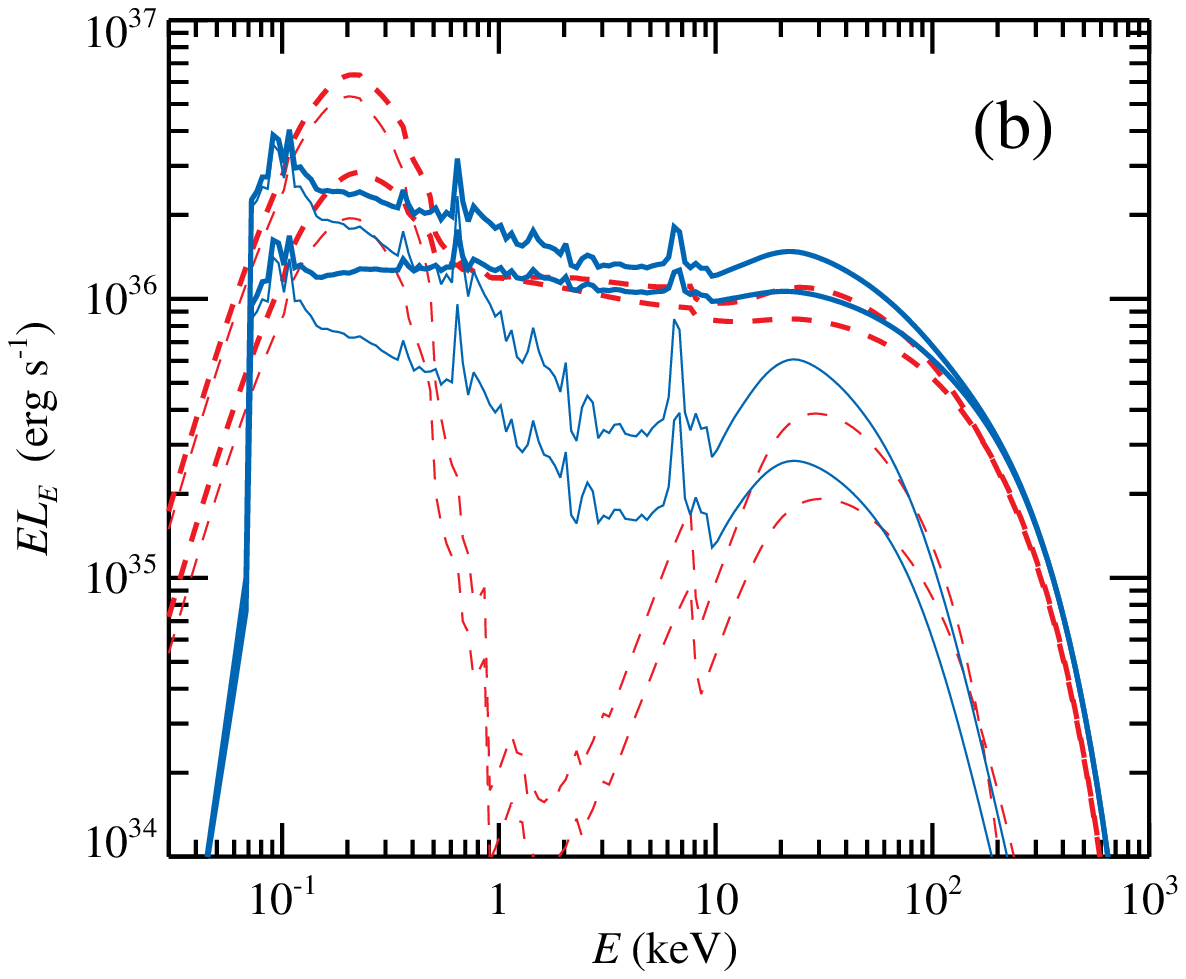}}
\caption{Comparison of the self-consistent spectra from the sandwich corona for two reflection models, {\sc xillverCp} (solid blue curves) and {\sc pexriv} (dashed red curves), for two different ionization parameters (a) $\log \xi=1$ and (b) $\log \xi=3$. 
The reflected and reprocessed blackbody spectra are shown by thin lines and the total spectrum by the thick lines.  
The upper and lower curves correspond to two inclinations, $i=18\degr$ and $60\degr$, respectively. 
} \label{fig:perxiv_vs_xillver} 
\end{figure*}

The presence of an additional source of seed photons not directly connected to the reflecting surface of the disc/clumps allows a certain freedom in the spectral modelling.
It easily resolves the photon starving problem.
Calculations of emitted spectra from hybrid electrons are, however, more complex compared to the case of fully thermal plasma,  as  not only the electron temperature, but also the whole distribution function for electrons 
has to be found self-consistently with the photon field they produce.
Hybrid models proved to give a good quantitative description for the X-ray/$\gamma$-ray spectra of both black hole binaries and active galactic nuclei  \citep{PV09,MB09,VVP11}, and at longer wavelengths \citep{VPV13,PV14}.

We simulate the spectrum of the inner parts of hot accretion flow, which is approximated by a sphere with radius $R=3\times 10^7$~cm, corresponding to 10 Schwarzschild radii for a $10\msun$ black hole, of the Thomson optical depth of $\tau_{\rm T}=1$.
Particle and photon distributions are assumed to be homogeneous and isotropic. 
The magnetic field is assumed to be tangled and characterized by its strength of $B=5\times 10^5$~G.
We assume that the energization proceeds through the injection of electrons with power-law energy distribution, $\dot n_{\rm e}(\gamma) \propto \gamma^{-\Gamma_{\rm inj}}$; we take $\Gamma_{\rm inj}=2.5$.
The injected electron distribution evolves because of the cooling and heating by cyclo-synchrotron, Compton scattering, bremsstrahlung and pair production, and by electron-electron Coulomb collisions.
The same number of electrons is removed from the evolved distribution to keep the Thomson optical depth constant.
The injected energy is regulated through the total amount of injected particles related to the total emitted luminosity of $L=10^{37}$~erg~s$^{-1}$.
The spectrum is calculated by solving a system of coupled relativistic kinetic equations for photons, electrons, and positrons.
Further details of modelling can be found in \citet{VP09}.

The resulting spectrum is shown in Fig.~\ref{fig:hybrid}. 
The 2--10~keV spectral index is $\Gamma=1.76$ and the equilibrium temperature (of the Maxwellian part of electron distribution) is $kT_{\rm e}=53$~keV.
The dotted curve shows the continuum that is produced by Comptonization of the synchrotron radiation from the hot flow.
An additional component related to the reflection and reprocessing of the Comptonized radiation in the outer cold disc is also added (dashed curve). In this case, we do not consider mutual interaction between reflected/reprocessed radiation and the hot flow because these components are likely physically separated. 
As was shown in \citet{PV09} and \citet{VPV13}, the hybrid thermal/non-thermal model predicts rather stable slopes of the Comptonized continuum if the total optical depth of the hot flow is of the order unity. 
The model is thus consistent with observations. 

\section{Discussion}

\subsection{Comparison of reflection models}
\label{sec:comp_refl} 

Many previous models of the X-ray emission from the accreting black holes used the  {\sc pexrav}/{\sc pexriv} model \citep{MZ95}. 
This is the simplest reflection model that correctly represents the angular-distribution of the reflected radiation using exact Klein--Nishina cross-section. 
It provides analytical formulae for the Green function for Compton reflection based on a Monte Carlo simulation of photon propagation in a homogeneous slab with an isotropic source above it. 
The reflected spectrum is computed as a convolution of the Green function with any incident spectrum (not only cut-off power law).
The {\sc pexrav} assumes neutral matter, while the {\sc pexriv} model is for the ionized material. 
The latter model ignores radiative transfer effects for the computation of the ionization/recombination balance \citep{Done92}. 

Radiative transfer effects have been  accounted for  by various authors (e.g. \citet{RF93}, \citet{GK10}, and \citet{GK13}).
For our analysis here we use the {\sc xillverCp} model, which is provided as a tabulated set of reflection spectra for a family of the incident Comptonization continua described by the approximate {\sc nthcomp} model. 
To test the role of the reflection model and the effect of the Comptonization model we ran two test cases with $\log\xi=1$  and $3$ and computed self-consistent sandwich-corona spectra using {\sc pexriv} and {\sc xillverCp} reflection models. 

In the first case ($\log\xi=1$), the greatest difference is observed in the reflection spectrum above 30~keV (see Fig.~\ref{fig:perxiv_vs_xillver}a). 
The most likely reasons being that the Klein--Nishina effects are not accounted for in {\sc xillverCp} and that the incident photons are injected at 45\degr\ to the slab normal, whereas {\sc pexriv} assumes an isotropic source. 
The difference at energies below 10~keV is likely caused by the choice of photoelectric cross sections (see also fig.~1 in \citealt{NZS16}).
However, the equilibrium coronal electron temperature differs by less than 1\% and the total spectra are nearly identical up to 20~keV. 
Thus the self-consistent spectrum does not depend significantly on the choice of the reflection or Comptonization model.  
At higher $\xi$ (see Fig.~\ref{fig:perxiv_vs_xillver}b), the {\sc xillverCp} reflected spectrum has substantially more photons in the 1--10~keV range and 
fewer photons below 1~keV than the {\sc pexriv} model.
However, this difference in the reflection spectrum only results  in a 5\% higher electron temperature and somewhat harder self-consistent Comptonization spectra by $\Delta\Gamma=0.18$.

One  of the major drawbacks of the considered reflection models  is the assumption of the constant density matter. 
A better approximation would be a constant-pressure material  \citep{MDM05}. 
In reality, of course, the disc atmosphere is in hydrostatic equilibrium with both density and pressure decreasing towards the surface. 
The X-ray illumination also  changes the atmosphere structure causing the top layers to expand, further lowering their density. 
Thermal ionization instability can then  lead to a rapid change in the ionization state of the material. 
The top layers are heated to nearly Compton temperature determined by the balance between Compton scattering heating and cooling and bremsstrahlung emission. 
These layers are almost completely ionized, while the layers below are much denser and nearly neutral. 
The reflection models that take into account all these effects  \citep{NKK00,NK01,DN01ApJ} are much more time consuming to compute. 
This makes them less flexible and limits their application to the data \citep{Barrio03,DG06}. 
The hydrostatic-equilibrium models, however, are not much different from the constant-density model at low ionization $\log\xi<2$ because in that case the gas is nearly neutral \citep{NKK00}. 
At higher ionization, the approximation of the constant-density slab used by {\sc xillverCp} predicts too few soft photons compared to a more physically realistic case of reflection from a constant-pressure slab \citep[see fig.~5 in][]{MDM05} or a slab in hydrostatic equilibrium \citep[see fig.~8 in][]{NKK00}. 
Thus, we conclude here that the results obtained using the {\sc xillverCp} reflection model are robust for low-ionization discs, but at high ionizations they predict  spectra that are too hard.

\subsection{Comparison to previous sandwich-corona models}
\label{sec:comp_sandwich}

Energy balance in the slab-corona geometry was first studied by \citet{HM91,HM93}, but their finding was that the resulting spectra are softer than those of the hard state of Galactic BH XRBs. 
As we discussed above, this conclusion was confirmed in a number of subsequent papers, as it is  here. 

On the other hand, it was proposed that the hard-state X-ray spectra in this geometry can be recovered by accounting for the reflected emission of the cold disc, which is then Compton up-scattered again as it passes through the hot medium (e.g. \citealt{YFRT01} and more recently \citealt{steiner17}). 
In the latter paper, it was argued that the truncated disc/hot flow geometry is not capable of reproducing the luminosities observed in the bright hard state of GX~339--4 for large truncation radii as this would require super-Eddington mass accretion rates. 
We analyse these arguments and compare their findings to our results.

The following set-up is adopted in \citet{steiner17}. 
The optically thick accretion disc in the hard-state accretion flow is described by the model of \citet{SS73}, but with the truncation radius being a free parameter, constrained to $r_{\rm in}\geq r_{\rm ISCO}$, and with the zero-stress boundary condition imposed at $r_{\rm in}$ (as implemented in the spectral-fitting routine of \citealt{ZNM05}). 
Then, the disc luminosity from the viscous dissipation from both sides of the disc is (in the Newtonian approximation)
\begin{equation}
L_{\rm intr}=\frac{\dot M_{\rm d} c^2}{2 r_{\rm in}/R_{\rm g}},
\label{bb_visc}
\end{equation}
implying that the accretion rate required to account for a given disc luminosity is $\dot M_{\rm d}\propto r_{\rm in}$, where $\dot M_{\rm d}$ is the mass flow rate through the disc.

The authors then consider Compton scattering of a fraction of the disc blackbody photons in the corona requiring conservation of photons, and using either a simple description of Comptonization as an e-folded power law or the thermal Comptonization model of \citet{ZJM96}. 
The Comptonized spectrum is emitted partly outside and partly towards the disc, which then gives rise to the reflection component,  using the model of \citet{GK10}. 
We denote the total Comptonization luminosity emitted both outside and towards the disc as $L_{\rm C}$, and the fraction of photons going towards the disc as $R_{\rm F}$.
The reflection component is smeared relativistically \citep{Dauser10}, and then Compton scattered in the corona, which is a major feature of the model.
As we noted above, many theoretical self-consistent models of hot plasma interacting with cold gas (see \citealt{SBS95,PS96,MBP01}) have accounted for scattering of the reflected radiation. 
However, this is not accounted for in the {\sc xspec} models like {\sc compps}.
The authors argue that the inclusion of Compton scattering of the reflected photons allows us to increase the relative strength of the fitted reflected component to values expected in the geometry of the corona lying above the disc. 
The value of reflection fraction, $R_{\rm F}$, fitted to the GX~339--4 data used in this study is $\simeq 0.8$. 

In the approach by \citet{steiner17} the seed photons for Compton scattering are dominated by the disc blackbody photons generated by viscous dissipation within the disc, which implies that $L_{\rm C}\propto L_{\rm intr}$.
At the same time a large truncation radius implies a large accretion rate in order to generate enough blackbody photons far away from the black hole, with the relationship of $\dot M_{\rm d}\propto r_{\rm in}$. 
This argument is very similar to the well-known photon starvation problem for the truncated disc/hot flow geometry.
However, we point out that  this problem appears only if we assume that the seed blackbody luminosity originates exclusively from  viscous dissipation. 
In the considered slab-corona  geometry, the disc is irradiated by the coronal emission, which in the hard state of BH XRBs has a much higher luminosity, typically  one order of magnitude, than that of the disc. 
\citet{steiner17} take into account the reflection from the disc, with the reflection fraction of $\simeq 0.8$, i.e. with similar fluxes going to the disc and outside in their best-fit model for GX 339--4. 
The typical albedo for the reflection in medium/hard X-rays is $a \simeq 0.2$--0.3, and thus most of the irradiating flux is actually absorbed and then re-emitted as soft photons, with only a relatively small fraction of it Compton-scattered within the disc and emitted outside,  resulting in $L_{\rm irr}\gg L_{\rm intr}$. 
Therefore, the disc will emit not only the viscously dissipated power (given by $\dot M_{\rm d}$), but also the power from the irradiation by the corona, which is  linked to the reflection fraction (see e.g. \citealt{GDP08}). 
Therefore, $\dot M_{\rm d}$ does not need to be high for a sufficient emission of blackbody photons.

Observationally,  it has been found that the disc component in the spectra of BH transients in the hard state does not  follow the disc-blackbody dependence of the luminosity on the inner radius (see e.g. \citet{GDP08} and \citet{BZ16}), which confirms the dominant role of the disc irradiation. 
Irradiation is important  in the inner parts of the disc and also in the outer parts. 
It has also been a general feature of accretion disc models since \citet{SS73} and it is crucial to explain the light 
curves, optical fluxes, and recurrence of BH transient systems (e.g. \citet{vP96}, \citet{KKB96}, \citet{GDP09}, \citet{CFD12} and \citet{PVR14}). 
In particular, irradiation reduces the critical luminosity below which BH or neutron-star LMXBs are transient by an order of magnitude. 

Our results (Fig.~\ref{fig:slab}) show that for a sandwich-corona geometry with zero viscous dissipation the spectra are soft.
This should be contrasted with the model of the hard state of GX 339--4 presented in fig.~4 and table~1 in \citet{steiner17}. 
There both high reflection and  high scattering fractions imply a geometry close to the slab corona, but the resulting spectrum is hard. 
The reason for this discrepancy is our self-consistent calculation of the electron temperature of the corona and the resulting Comptonization spectra, omitted by \citet{steiner17}. 
Their treatment of reflection also has multiple problems. 
First, the photon energy grid in the {\sc xillver} and {\sc relxill} models only extends down to 70 eV, as discussed in Sect.~\ref{sec:sandwich}, but the number of photons in the reflected spectrum below that energy exceeds that above 70 eV by orders of magnitude (see fig.~3 in \citealt{GK13}). 
This then implies that the normalization of the Comptonization continuum relative to the reflection spectrum is wrong  by orders of magnitude and that Comptonization  does not conserve photons. 
Furthermore, the limited photon energy range also leads to the angle-integrated reflected luminosity being different from the illuminating luminosity, i.e. the energy in these models is not conserved either. 
And finally, the angular distribution of reflected and Comptonized photons is different and therefore it is meaningless to demand the photon conservation for the spectra observed at one specific inclination.  

We note here that spectra obtained for the geometries where the hot flow contains cold clumps or is overlapping with the disc (Fig.~\ref{fig:clumps}) can be made hard, similar to the observed ones. 
The problem of requiring a super-Eddington accretion rate for the truncated disc geometry disappears if we take into  account  soft photons from other sources, such as  reprocessing and internal synchrotron emission. 

\subsection{Preferred geometry}

As we argued above, sandwich, patchy, and outflowing corona models need a lot of fine-tuning to reproduce commonly observed hard  spectra and electron temperatures.   
Moreover, the models assuming the disc is going down to the ISCO are inconsistent with the spectral modelling of reflection and the K${\alpha}$ line \citep{GCR00,PFP15,BZ16}, which suggest $r_{\rm in}\sim 100 R_{\rm g}$ in the hard state.
The model where the disc does not overlap with the hot flow is yet another extreme, expected to have too hard spectra.
Three other models (hot flow with cold clumps, hot flow overlapping with the cold disc, and hot flow with cyclo-synchrotron photons) are degenerate in the X-ray spectral data alone.
However, they are expected to make different predictions for the X-ray short- and long-term variability, optical/infrared (OIR) spectra, and interconnection of emission at different wavelengths.
Many discoveries in terms of multiwavelength behaviour were made after those geometries were proposed, and properties for some of them were not discussed in application to the new data.
We now summarize the timing and multiwavelength properties critical to discriminate between the geometries.

\subsubsection{X-ray timing properties}

Significant hard time lags between different X-ray energy bands were detected, with a peculiar power-law dependence on frequency \citep{MKM88,Nowak99a} and a logarithmic dependence on the energy band \citep{KCG01}.
This behaviour was interpreted both in terms of flares with pivoting power law \citep{PF99}, corresponding to a patchy corona geometry, and in terms of radial dependence of the emitted spectrum \citep{KCG01}, which is  expected in the truncated disc/hot flow geometry.
The observed power-law dependence of the X-ray variability power on Fourier frequency \citep[e.g.][]{Nowak99a} and the linear dependence of root mean square variability amplitude on flux \citep{UM01} argue against the model where X-rays are produced in the flares, and support the picture of accretion rate fluctuations propagating in the radially stratified hot medium \citep{Lyub97,KCG01}. 
The inferred radial dependence of the X-ray spectral hardness, with harder spectra coming from the innermost parts, naturally arises in the geometries with hot flow overlapping with the disc and in the hot flow with non-thermal particles.
On the other hand, this dependence is not expected to be automatically satisfied in the case of clumpy hot flow.

Another powerful technique used to probe the accretion flow geometry is  X-ray reverberation \citep{RYBF99,Pou02,UCFK14}. 
Measurements of the time delays between X-ray continuum and the signatures of the cold accretion disc, namely a soft thermal component and iron K${\alpha}$ line \citep{DeMarco15reverber,demarco17,DeMarco16reverber}, suggest the truncated disc geometry, where the  truncation radius decreases with rising luminosity, i.e. towards the soft state.

An independent line of arguments for a truncated disc geometry comes from the studies of the equivalent width of the iron K${\alpha}$ line in a range of Fourier frequencies \citep{GCR00}.
A substantial suppression of the iron line variability at high Fourier frequencies corresponding to the innermost parts of the accretion flow was found, which implies that  the cold disc truncation radius is around  $\sim100 R_{\rm g}$  in the hard state.

These findings disfavour clumpy accretion flow geometry, unless these clumps are fully ionized and are acting as mirrors from the spectral point of view.
On the other hand, such high values for the truncation radius results in small number of seed photons entering the zone of major energy release (within $\sim10 R_{\rm g}$), leading to the standard problem with photon starvation. 
Thus, high truncation radii in the hard state disfavour the models where seed photons  come solely from the disc.

\subsubsection{Multiwavelength properties}

We  now discuss various possibilities involving truncation of the disc.
The major question here concerns the source of seed photons, either thermal (viscous or reprocessed) photons from clumps/accretion disc or synchrotron photons.
At the outburst peak, close to the soft state, the accretion disc is likely to be the major source of seed photons.
As the outburst declines, the spectra are found to harden, and the trend is reversed at very low luminosities 
\citep[below $\sim10^{-3}L_{\rm Edd}$,][]{SPDM11}, i.e. the spectra are softer for lower luminosities.
Spectral hardening is expected from the outward movement of the disc inner radius; however, the softening is not, and it instead requires an additional source of synchrotron seed photons \citep{VVP11}. 
The softening can be explained in the clumpy hot flow scenario if ionization drops as the luminosity decreases. 
It is difficult to explain the constant hardness ratio during the rising phase (when the observed luminosity is in the range $\sim 10^{-3}-10^{-2} L_{\rm Edd}$) using the geometry where seed photons come solely from the disc overlapping with the hot flow or from the cold clumps.
The synchrotron self-Compton mechanism is preferred here. 
The switch of the source of seed photons likely occurs close to the point where the spectra start to soften on the rising phase, or achieve the hardest spectra on the decline stage.
The critical luminosity (of about 2\% $L_{\rm Edd}$) where these two seed photon populations switch has  recently been identified at the decline stage of SWIFT~J1753.5--0127 \citep{KVT16}.

Evolution of the X-ray spectrum towards and from the soft states proceeds through the intermediate states when the spectra are flat ($\Gamma\sim 2$). 
These spectra are not well fitted with one Comptonization continuum, and  two continua are required instead \citep{IPG05,SUT11}.
It is then natural to interpret these continua as being produced in two physically separated regions with different sources of  seed photons: disc and synchrotron.  
These two continua are expected to change in response to fluctuations in mass accretion rate, one delayed with respect to another, producing an interference picture in the power spectrum, which we see in the data \citep{V16}.

In the hard state, the scenario with the synchrotron seed photons predicts that OIR radiation is partially produced by the synchrotron component from the hot accretion flow, thus properties of emission at longer wavelengths can serve as a diagnostics of accretion geometry.
The OIR excesses above the possible accretion disc fluxes were reported in a number of sources \citep{CHM03,CDSG10,GBRC11}.
An excess flux at these wavelengths can be produced by the jet \citep{BK79}; however, in some sources either the extrapolation of the power law from radio significantly underpredicts OIR fluxes or the extrapolated slope is different \citep{CHM03,DGS09}.
The jet scenario also has major difficulties in explaining the optical/X-ray cross-correlation function with broad anti-correlation observed in SWIFT~J1753.5--0127 \citep{DGS08,DSG11}, which perhaps possesses the faintest radio jet.
On the other hand, this shape of the cross-correlation function is expected to be seen if the hot accretion flow contributes to both optical and X-ray energy bands \citep{VPV11}. 
The observed evolution of the shape of the cross-correlation function, from demonstrating a positive correlation at the outburst peak to the anti-correlation feature in the outburst tail, is consistent with the scenario where the cold disc recedes and is replaced by the hot flow \citep{VGH17}.
Neither high OIR fluxes, nor short-term variability in the hard state can be understood in terms of the scenario of clumps/intersecting disc.

The long-term multiwavelength behaviour can also be used to constrain the accretion geometry. 
The outburst X-ray light-curve resembles a fast rise exponential decay profile, and the light-curves in OIR energy bands often demonstrate additional prominent flares lasting a substantial fraction of the outburst duration.
The origin of these flares has been discussed in a context of hot flow and jet synchrotron emission \citep{Buxton12,PVR14}.
The jet hypothesis is inconsistent with the spectrum of the flare at least in XTE J1550--564, which was shown to be very hard when the flare starts \citep{PVR14}.
On the other hand, the evolution of the flare spectrum is in good agreement with the hot flow scenario.
Neither a cold accretion disc nor emission of cold clumps is able to account for the appearance of these flares, which favours the scenario of hot flow with cyclo-synchrotron radiation in the hard state.

\section{Summary}

We have modelled Comptonization and photon reprocessing processes in various geometries relevant to the hard state of BH XRBs. We have used a modern (constant-density) reflection code, {\sc xillverCp}, to account for Compton reflection and reprocessing. Our main results are as follows.

Our self-consistent treatment confirms earlier studies, in particular that a static disc-corona is not a viable geometry for the hard state of BH X-ray XRBs, even accounting for scattering of reflection radiation in the corona. 
In the absence of intrinsic disc dissipation (energy is dissipated entirely in the corona), the sandwich-corona models require a high ionization parameter ($\log\xi\gtrsim3.5$) to agree with the observed spectral slopes. 
More physically realistic reflection models that account for density gradients and allow for thermal instability to operate in the disc, predict spectra softer than the observed ones even at such high ionizations. 
Alternative models, such as patchy or outflowing corona and clumpy hot flow can reproduce the observed spectra and electron temperatures after fine-tuning  the parameters.
We show that the spectral slopes heavily depend on the fraction of energy dissipated intrinsically in the disc.
Assuming only a 10 \% energy release proceeding through the viscous heating in the cold disc, we obtain  spectra that are too soft for the above-mentioned models.
Given that the energy liberation almost entirely proceeds through the viscous heating in the soft state, it is not clear why such a tight upper boundary should apply to the hard state.

We conclude that the disc has to be truncated before the ISCO, and that it is replaced by some kind of a hot flow. 
This assures an agreement with the X-ray observations of BH XRBs in the hard state, namely their spectral slopes, the range of electron temperatures, and the ionization levels of the cold matter. 
Some mixing of the cold and hot media, either the disc overlapping with the hot flow or the hot flow containing some cold medium, can reproduce the observed ranges of parameters but does not reproduce the observed evolution of these parameters over the course of the outburst (nearly constant hardness ratio at moderate $L/L_{\rm Edd}$ and spectral softening towards low luminosities). 
We also reiterate the importance of the cyclo-synchrotron emission by hybrid electrons (mostly thermal with a small non-thermal tail)  as the process providing both seed photons for Comptonization and low-energy (OIR) photons.
The latter mechanism is favoured by the multiwavelength spectra and variability properties.

\begin{acknowledgements}
We thank A. Beloborodov and J. Steiner for valuable discussions, and J. Garc{\'{\i}}a for help with the {\sc xillverCp} model. 
This research has been supported in part by the Polish National Science Centre grants 2013/10/M/ST9/00729, 2015/18/A/ST9/00746, and 2016/21/P/ST9/04025; the Academy of Finland grant 309308; and  grant 14.W03.31.0021 of the  Ministry of Education and Science of the Russian Federation.
\end{acknowledgements}


\end{document}